\documentclass[runningheads]{llncs}
\usepackage{amsfonts}
\usepackage[dvipsnames]{xcolor}
\usepackage{booktabs}
\usepackage{amsmath}

\newcommand{\Xmat}{{\bf X}}
\newcommand{\Ymat}{{\bf Y}}
\newcommand{\Zmat}{{\bf Z}}

\usepackage{soul}

\newcommand{\EE}{\mathbb{E}}

\newcommand{\BR}{\mathbb{R}}
\newcommand{\PP}{\mathbb{P}}

\newcommand{\BS}{\mathbb{S}}

\usepackage{tikz}

\newcommand{\beq}{\begin{equation}}
\newcommand{\eeq}{\end{equation}}
\newcommand{\beqs}{\begin{eqnarray}}
\newcommand{\eeqs}{\end{eqnarray}}
\newcommand{\barr}{\begin{array}}
\newcommand{\earr}{\end{array}}
\usepackage{multibib}
\newcites{SM}{SM References}

\usepackage{bbm}
\usepackage[ruled]{algorithm2e}
\newcommand{\IE}{{\it i.e.}}
\newcommand{\EG}{{\it e.g.}}
\usepackage{wrapfig}
\usepackage{graphicx}
\newcommand{\beginsupplement}{%
        \setcounter{table}{0}
        \renewcommand{\thetable}{S\arabic{table}}%
        \setcounter{figure}{0}
        \renewcommand{\thefigure}{S\arabic{figure}}%
     }
\begin{document}
\title{Feedback Effect in User Interaction with Intelligent Assistants: Delayed Engagement, Adaption and Drop-out}
\author{Zidi Xiu\inst{1}\textsuperscript{$\dagger$}, Kai-Chen Cheng\inst{1}, David Q. Sun\inst{1}\textsuperscript{$\dagger$}, Jiannan Lu\inst{1}, Hadas Kotek\inst{1}, Yuhan Zhang\inst{2}\thanks{\small{Contributions made during the internship at Apple in the summer of 2022.}}, Paul McCarthy\inst{1},  Christopher Klein\inst{1}, Stephen Pulman\inst{1}, Jason D. Williams\inst{1}
}

\authorrunning{Z. Xiu, K. Cheng, D. Sun et al.}
\institute{Apple, One Apple Park Way, Cupertino, CA 95014, USA \and
Department of Linguistics, Harvard University, Cambridge, MA 02138\\
\textsuperscript{$\dagger$}\email{\{z\_xiu,dqs\}@apple.com}}
\titlerunning{The Feedback Effect with IA}
  
\maketitle              
\begin{abstract}
With the growing popularity of intelligent assistants (IAs), evaluating IA quality becomes an increasingly active field of research. 
This paper identifies and quantifies the \textit{feedback effect}, a novel component in IA-user interactions -- how the capabilities and limitations of the IA influence user behavior over time. 
First, we demonstrate that unhelpful responses from the IA cause users to delay or reduce subsequent interactions in the short term via an observational study. 
Next, we expand the time horizon to examine behavior changes and show that as users discover the limitations of the IA’s understanding and functional capabilities, they learn to adjust the scope and wording of their requests to increase the likelihood of receiving a helpful response from the IA. Our findings highlight the impact of the feedback effect at both the micro and meso levels. We further discuss its macro-level consequences: unsatisfactory interactions continuously reduce the likelihood and diversity of future user engagements in a feedback loop.
\vspace{-1em}
\keywords{Data Mining \and Intelligent Assistant Evaluation}
\end{abstract}
\vspace{-3em}
\section{Introduction}
\vspace{-1em}
Originated from spoken dialog systems (SDS), intelligent assistants (IAs) had rapid growth since the 1990s \cite{glass1999challenges}, with both research prototypes and industry applications. 
As their capabilities grow with recent advancements in machine learning and increased adoption of smart devices, IAs are becoming increasingly popular in daily life \cite{de2020intelligent,kepuska2018next}. Such IAs often offer a voice user interface, allowing users to fulfill everyday tasks, get answers to knowledge queries, or start casual social conversations, by simply speaking to their device \cite{lee2015natural,purington2017alexa};
that is, they take human voice as input, which they process in order to provide an appropriate response \cite{santos2016intelligent}. The evolution of these hands-free human-device interaction systems brings new challenges and opportunities to the data mining community. 


IA systems often consist of several interconnected modules: Automated Speech Recognition (ASR), Natural Language Understanding (NLU), Response Generation \& Text-to-Speech (TTS), Task Execution (\EG, sending emails, setting alarms and playing songs), and Question Answering \cite{glass1999challenges,jiang2015automatic,lopatovska2019talk}.
Many of the active developments in the field are formulated as supervised learning problems where the model predicts a target from an input, e.g., a piece of text from a speech audio input (ASR), a predefined language representation from a piece of text (NLU), or a clip of audio from a string of text (TTS). Naturally, the evaluation of these models often involves comparing model predictions to some ground-truth datasets. 

When building such an evaluation dataset from real-world usage, we inevitably introduce user behavior into the measurement. User interactions with IA are likely to be influenced by the their pre-existing perception of IA's capabilities and limitations, therefore introducing a bias in the distribution of ``chances of success'' in logged user interactions -- users are more likely to ask what they know the IA can handle. This hypothesis makes intuitive sense and has been partly suggested by an earlier study on vocabulary convergence in users learning to speak to an SDS \cite{levow2003learning}. 

In this context, we define \emph{feedback effect} as the behavior pattern changes in users of an interactive intelligent system (\EG, IA) that are attributable to their cumulative experiences with said system. Our contributions can be summarized as follows. First, we establish a causal link between IA performance and immediate subsequent user activities, and quantify its impact on users of a real-world IA. Second, we identify distinct dynamics of behavior change for a cohort of new users over a set period of time, demonstrating how users first explore the IA's capabilities before eventually adapting or quitting.
Third, having examined the \emph{feedback effect} and its impact in detail, we
provide generalizable recommendations to mitigate its bias in IA evaluation.

\vspace{-1em}
\section{Related Work}
\vspace{-1em}
 \textbf{IA evaluation methods and metrics.} Many studies have been devoted to addressing the challenges in IA evaluation. Objective metrics like accuracy cannot present a comprehensive view of the system \cite{gao2019neural}.
 Human annotation is a crucial part of the process, but it incurs a high expense and is hard to scale \cite{kiseleva2016understanding}. Apart from human evaluation, \IE, user self-reported scores or annotated scores, subjective metrics have been introduced. Jiang \cite{jiang2015automatic} designed a user satisfaction score prediction model based on user behavior patterns, ungrammatical sentence structures, and device features. Other implicit feedback from users (\EG, acoustic features) are helpful to approximate success rates \cite{komatani2007analyzing}.

\textbf{User adaptation and lexical convergence.} 
\emph{Adaptation} (or \emph{entrainment}) describes the phenomenon whereby the vocabulary and syntax used by speakers converge as they engage in a conversation over time
\cite{reitter2006computational}. Convergence can be measured by observing repetitive use of tokens in users' requests \cite{duplessis2017automatic} and high frequency words \cite{hirschberg2008high}. Adaptation happens subconsciously and leads to more successful conversations \cite{friedberg2012lexical}.
In SDS, the speakers in a dialogue are the IA and the user. When the IA actively adapts to the user in the conversation, the quality of the generated IA responses increases substantially \cite{walker2007individual,wen2015semantically}. The phenomenon of lexical adaptation of users to the IA system has been investigated as well \cite{levow2003learning,parent2010lexical}. Currently, most IAs are built upon a limited domain with restricted vocabulary size \cite{chattaraman2019should}. Users' vocabulary variability tends to decrease as they engage with the IA over time. 
This naturally limits the linguistic diversity of user queries, although out-of-domain queries can happen from time to time  \cite{glass1999challenges}. 


\vspace{-1.5em}
\section{Data Collection}
\vspace{-1em}
We analyzed logged interactions (both user queries and the associated IA responses) from a real-world IA system. All data originate from users who have given consent to data collection, storage, and review. The data is associated with a \emph{random, device-generated} identifier. Therefore, when we use the term `user' in the context of the interaction data analysis, we are \emph{actually} referring to this random identifier. While the identifier is a reasonable proxy of a user, we must recognize its limitations -- in our analysis, we are unable to differentiate multiple users who share a single device to interact with the IA, nor to associate requests from a single user that were initiated on multiple devices. 

The population of interest is US English-speaking smartphone users who interacted with the IA in 2021 and 2022. We randomly sampled interaction data from two distinct time periods \emph{before} and \emph{after} a special event in late 2021. This event entailed new software and hardware releases, potentially introducing nontrivial changes to user behavior and demographics, while simultaneously presenting unique opportunities for our particular investigation. 

\vspace{-1.5em}
\subsection{Study 1: Pre-event Control Period}\label{sec:causalData}
\vspace{-0.5em}
To investigate the feedback effect on user engagement, we randomly sampled interaction data from a two-week period in August 2021. The choice of a relatively \emph{short} time period \emph{before} the special event helps us (i) directly control for seasonality and (ii) avoid the impact of the special event, where product releases and feature announcements usually stimulate user engagement and attract new users. (We return to a discussion of new users below.)


We further control for software and hardware versions, before taking a random sample of approximately 14,000 users who had at least one interaction with the IA during the study period. We then randomly sampled \emph{one} interaction per user and used human-label review to determine whether the IA response was helpful. We additionally analyzed the frequency of interactions for the user in the 2 weeks prior to and 2 weeks following our causal analysis. 
 In our sample, approximately 80\% of the interactions were labeled as \emph{helpful} to the user.\footnote{\tiny{This value is not necessarily a reflection of the aggregated or expected satisfaction metric, due to the sampling method and potential bias in the subpopulation of choice.}} The results of this study are presented in Section \ref{sec:causal}.

\vspace{-1.3em}
\subsection{Study 2: Post-event New User Period}\label{sec:collapseData}
\vspace{-0.5em}

To investigate language convergence among a new user cohort, we randomly sampled data from a six-month period immediately \emph{after} the special event. With our interest in analyzing long-term behavior changes of a new user cohort, this choice of sampling period has two interesting implications. First, special events often lead to a surge in new users of the IA. Second, feature announcements at the special event may cause some existing users to perceive the updated IA as ``new" and explore it with a mindset akin to that of a new user. Given the challenges inherent to determining new user cohorts (to be further discussed in Section~\ref{sec:semanticConvergence}), these two factors are valuable as they collectively increase the share of new users, thus boosting the observability of the cohort. From this six-month period, we took a random sample of 5,000  users who used the new software version of the IA. For each user, all interactions with the IA in the full study period were used in our analysis. 
The study is described in Section \ref{sec:semanticConvergence}. 

\vspace{-1.5em}
\section{Feedback Effect on Engagement}\label{sec:causal}
\vspace{-1.0em}

\begin{wrapfigure}{l}{0.42\linewidth}
\vspace{-1.8em}
    \centering
    \includegraphics[width=\linewidth]{ 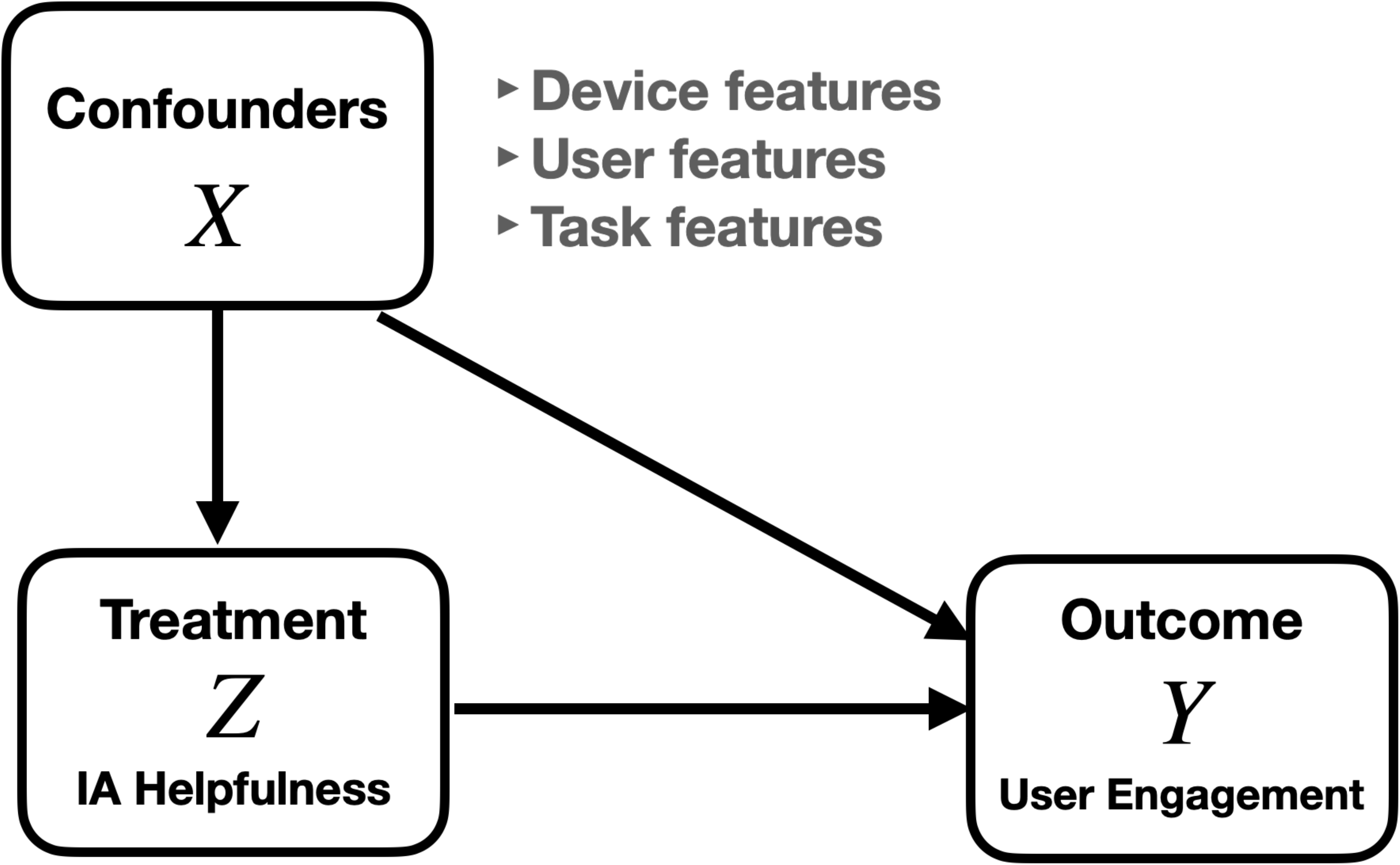}
    \caption{Causal graph illustrating the observational study of IA feedback effect accounting for the existence of confounding factors. 
    }
    \label{fig:causalDiag}
    \vspace{-2em}
\end{wrapfigure}
Intuitively, unhelpful responses from an IA may discourage users from future interactions with the IA. Our work aims to empirically shed light on the relation between IA helpfulness\footnote{\tiny{The IA helpfulness of a given user request is defined as the user's satisfaction with the IA's response to the request, as determined by human annotators. 
}} (\emph{helpful} or \emph{unhelpful} on a single interaction) and users' subsequent engagement patterns with the IA, as illustrated in Figure~\ref{fig:causalDiag}. To establish such a causal relationship from IA performance to user engagement, we adopt an observational analysis framework with IA related features.



Given a dataset with $N$ users, we denote unit $i$ having (i) covariates $\Xmat_i \in \BR^p$, (ii) a treatment variable $Z_i \in \{0,1\}$, indicating users experienced an \emph{unhelpful} interaction with the IA or a \emph{helpful} one respectively, and (iii) what would have happened if the unit is assigned to treatment and control, denoted by $Y_i(1)$ and $Y_i(0)$ respectively, according to the potential outcomes framework \cite{holland1986statistics,rubin1974estimating}. Consequently, the causal effect for unit $i$ is defined as $\tau_i = Y_i(1) - Y_i(0)$, namely the difference between the outcomes if treated differently on the same user. However, the fundamental problem of causal inference is that only the potential outcome -- the outcome in the group the subject was assigned to -- can be observed, \IE, $Y_i = Z_iY(Z_i) + (1-Z_i)Y(1-Z_i).$ Individual level causal estimands, the contrast of values between the two potential outcomes, cannot be expressed with functions of observed data alone. Consequently, our primary focus is on population level causal effects like the Average Treatment Effect (ATE),
$\tau = \EE ( \tau_i )$.


With a randomized controlled experiment, treated assignment mechanism is known and unconfounded, therefore we can directly and  accurately estimate and infer causal effects (\EG, ATE) from the observed data. However, in real-world scenarios, delicately designed experiments can be difficult or impossible to conduct. Instead, we must rely on observational techniques.

\vspace{-1.5em}
\subsection{Covariates and Outcome Variables}\label{sec:confounding}
\vspace{-1em}

Observational studies are susceptible to \textit{selection bias} due to confounding factors, which affect both the treatment $\Zmat$ and the outcome $\Ymat$, as
shown in Figure~\ref{fig:causalDiag}. 
To address any confoundness to the best of our ability, we have collected rich sets of the following IA related features, and assuming that there are no unobserved observed confounders.
(i) Device features: The type of device used to interact with the IA system, and the operating system version, (ii) Task features: The input sentence transcribed by the ASR system, the number of tokens in the input sentence, the word error rate (WER) of the transcription, and the domain that the IA executed with a confidence score provided by the NLU model (\EG, weather, phone, etc.), (iii) User related features: Prior activity levels measured as the number of active days before the interaction, and temporal features including local day of the week and time of the day when the interaction happened.

To quantify user engagement after the annotated IA interaction, Section \ref{sec:time2event} focuses on time to next session (``immediate shock"), and Section \ref{sec:NactiveDays} focuses on active day counts (``aftermath"). 

\vspace{-1.5em}
\subsection{Observational Causal Methods}
\vspace{-0.8em}

\indent\textbf{Matching methods.} Matching is a non-parametric method to alleviate the effects of confounding factors in observational studies. The goal is to obtain well-matched samples from different treatment groups, hoping to replicate randomized trial settings. The popular Coarsened Exact Matching (CEM) model is based on a monotonic imbalance reducing matching method at a pre-defined granularity with no assumption on assignment mechanisms \cite{iacus2012causal}. 

\textbf{Weighting methods.} 
Apart from matching covariates, weighting based methods use all of the high-dimensional data via summarizing scores, like the propensity score (PS).
PS reflects the probability of being assigned to treatment based on user's background attributes \cite{li2018balancing,rubin1974estimating}, $e(x)=P(Z_i=1|X_i=x)=\EE(\Zmat|\Xmat)$
Since the true PS is unknown, we adopt generalized linear regression models (GLMs) to estimate it, which are widely adopted by the scientific community.

With PS estimates available, the next question is how to leverage them. Li \cite{li2018balancing} proposed a family of balancing weights which enjoys balanced weighted distributions of covariates among treatment groups. Inverse-Probability Weights (IPW) are a special case of this family, shown in Eq.\eqref{eq:IPW}. 
As the name suggests, the weight is the inverse of the probability that a unit is assigned to the observed group, and the corresponding estimand is the ATE. However, IPW is very sensitive to outliers, \IE, when PS scores approach 0 or 1. 
To mitigate this challenge, Overlap Weights (OW) which emphasize a target population with the most covariate overlap \cite{li2018balancing}, shown in Eq.\eqref{eq:OverlapWeight}. 

\begin{minipage}{0.49\linewidth}
\begin{equation}\label{eq:IPW}
            \begin{cases}     w^{\text{IPW}}_1(x) =\frac{1}{e(x)}\\   w^{\text{IPW}}_0(x) =\frac{1}{1-e(x)} \end{cases}
\end{equation}
\end{minipage}
\begin{minipage}{0.49\linewidth}
\begin{equation}\label{eq:OverlapWeight}
        \begin{cases}     w^{\text{OW}}_1(x) =(1-e(x))\\      w^{\text{OW}}_0(x) =e(x), \end{cases}
\end{equation}
\end{minipage}
\noindent where $w_1$ corresponds to the weight assigned to the treatment group, and $w_0$ to the control group, respectively. Then the population level causal estimands of interest, the Weighted Average Treatment Effect (WATE), are derived from the balanced weights.
The target population varies with different weighting strategy.
The causal estimand shown in Eq. \eqref{eq:WATE} then becomes the average treatment effect for the overlap population.
\begin{equation}\label{eq:WATE}
    \hat{\tau}^w=\frac{\sum_{i=1}^Nw_1(\bf{x_i})Z_iY_i}{\sum_{i=1}^Nw_1(\bf{x_i})Z_i} - \frac{\sum_{i=1}^Nw_0(\bf{x_i})(1-Z_i)Y_i}{\sum_{i=1}^Nw_0(\bf{x_i})(1-Z_i)}
\end{equation}


\begin{figure}[t!]
    \centering
    \includegraphics[width=.65\linewidth]{ 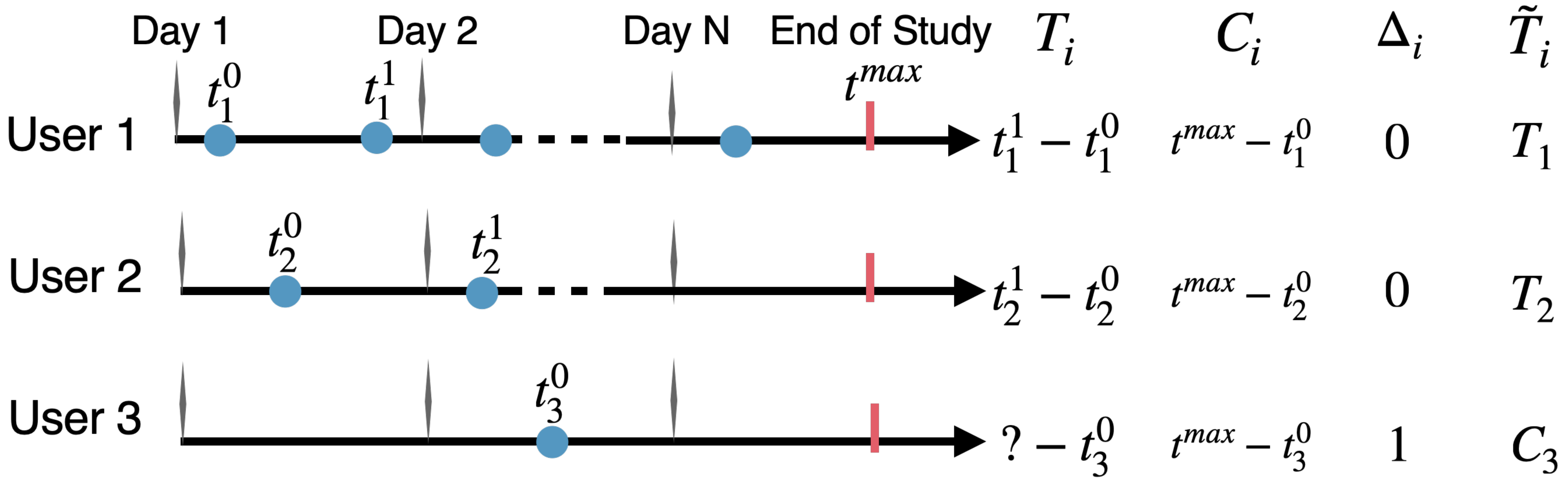}
    \caption{Illustration of the users' engagement (blue dots) after the request was annotated at time $t_i^0$ for user $i$. We observed the time-to-next-engagement for user 1 and 2, but user 3 was censored (the next engagement was not observed).}
    \label{fig:time2eventIllustration}
    \vspace{-2.2em}
\end{figure}

\vspace{-2em}

\subsection{Time to Next Engagement }\label{sec:time2event}
\vspace{-0.7em}

In this section, we establish causal links between interaction quality with the IA (as implied by the annotated helpfulness) 
and the user's time to next engagement. Specifically, our main hypothesis is that if a user has a helpful interaction with the IA, they are more likely to further engage with the IA in the future.

Unlike standard observational studies with well-defined and observable outcomes, time-to-event measures fall into the range of survival analyses, which focus on the length of time until the occurrence of a well-defined outcome \cite{miller2011survival,xiu2020variational}. 
A characteristic feature in the study of time-to-event distributions is the presence of \emph{censored} instances: events that do \emph{not} occur during the follow-up period of a subject. This can happen when the unit drops out during the study (right censoring), presenting challenges to standard statistical analysis tools. 

As illustrated in Figure~\ref{fig:time2eventIllustration}, assume for user $i$: the time-to-next engagement is $T_i$ with censoring time $C_i$, the observed outcome $\tilde{T}_i=T_i\wedge C_i$, and the censoring indicator $\Delta_i=\mathbbm{1}\{T_i\le C_i\}$. Under time-to-event settings, we observe a quadruplet $\{Z_i, \Xmat_i, \tilde{T}_i, \Delta_i\}$ for each sample. 
Each user also has a set of potential outcomes, $\{T_i(1), T_i(0)\}$.
Users may use the IA system at some point in our research and be assigned a \emph{helpfulness} score, but not show up again before the data collection period ends (\EG, User $3$ illustrated in Figure~\ref{fig:time2eventIllustration}). This yields a censored time $C_3$ instead of a definite time-to-next-engagement outcome $T_3$ which is not observed within the study period.

Following Zeng \cite{zeng2021propensity}, the causal estimand of interest is defined based on a function of the potential survival times, $\nu(T_i(z);t)=\mathbbm{1}\{T_i(z)\ge t\}$. It can be interpreted as an at-risk function with the potential outcome $T_i(z)$. The expectation of the risk function corresponding to the potential survival function of user $i$, \IE, the probability of no interaction with the IA until time $t$. Accordingly, the \emph{re-engagement} probability for users in treatment group $z$ within time $t$ is therefore defined as Eq.\eqref{eq:SPCE}.

 \begin{minipage}{0.55\linewidth}
\vspace{-10pt}
\begin{equation}\label{eq:SPCE}
    \EE[\nu(T_i(z);t)]=\PP[T_i(z)\ge t]=\BS_i(t;z)
\end{equation}
\end{minipage}
\begin{minipage}{0.40\linewidth}
 \begin{equation}\label{eq:prob-engage}
    \PP(t;z)=1-\BS(t;z)
 \end{equation}
\end{minipage}

To properly apply balancing weights \eqref{eq:OverlapWeight} with survival outcomes, right censoring needs to be accounted for. Pseudo-observation is therefore constructed based on re-sampling (a jack-knife statistic) and is interpreted as the individual contribution to the target estimate from a complete sample without censoring \cite{andersen2017causal}. Given a time $t$, denote the expectation of the risk function at that time point , \IE, $\EE[\nu(T_i(z);t)]$ in Eq.\eqref{eq:SPCE}, as $\theta(t)$,
which is a population parameter. Without loss of generality, we discuss the pseudo observation omitting the potential outcome notations.
The pseudo-observation for each unit $i$ can be specified as,
    $\hat{\theta}_i(t) = N\hat{\theta}(t) - (N-1)\hat{\theta}_{-i}(t)$,
where $\hat{\theta}(t)$ is the Kaplan-Meier estimator of the population risk at time $t$, which is based on $\Delta_i$ and $T_i$. $\hat{\theta}_{-i}(t)$ is calculated without unit $i$.
In this way, classic propensity score methods become applicable. 
Then the conditional causal effect averaged over a target population at time $t$ is:
\begin{align}
\hat{\tau}^w(t)&=\frac{\sum_{i=1}^Nw_1(\bf{x_i})Z_i \hat{\theta}_i(t)}{\sum_{i=1}^Nw_1(\bf{x_i})Z_i} - \frac{\sum_{i=1}^Nw_0(\bf{x_i})(1-Z_i) \hat{\theta}_i(t)}{\sum_{i=1}^Nw_0(\bf{x_i})(1-Z_i)} \nonumber\\ 
    &=(1-\hat{\BS}^{w_0}(t;0)) - (1-\hat{\BS}^{w_1}(t;1))= \hat{\PP}^{w_0}(t;0) - \hat{\PP}^{w_1}(t;1) \label{eq:pseudo-survival}
\end{align}

\begin{figure}[t]
\vspace{-1.5em}
    \centering
    \includegraphics[width=.7\linewidth]{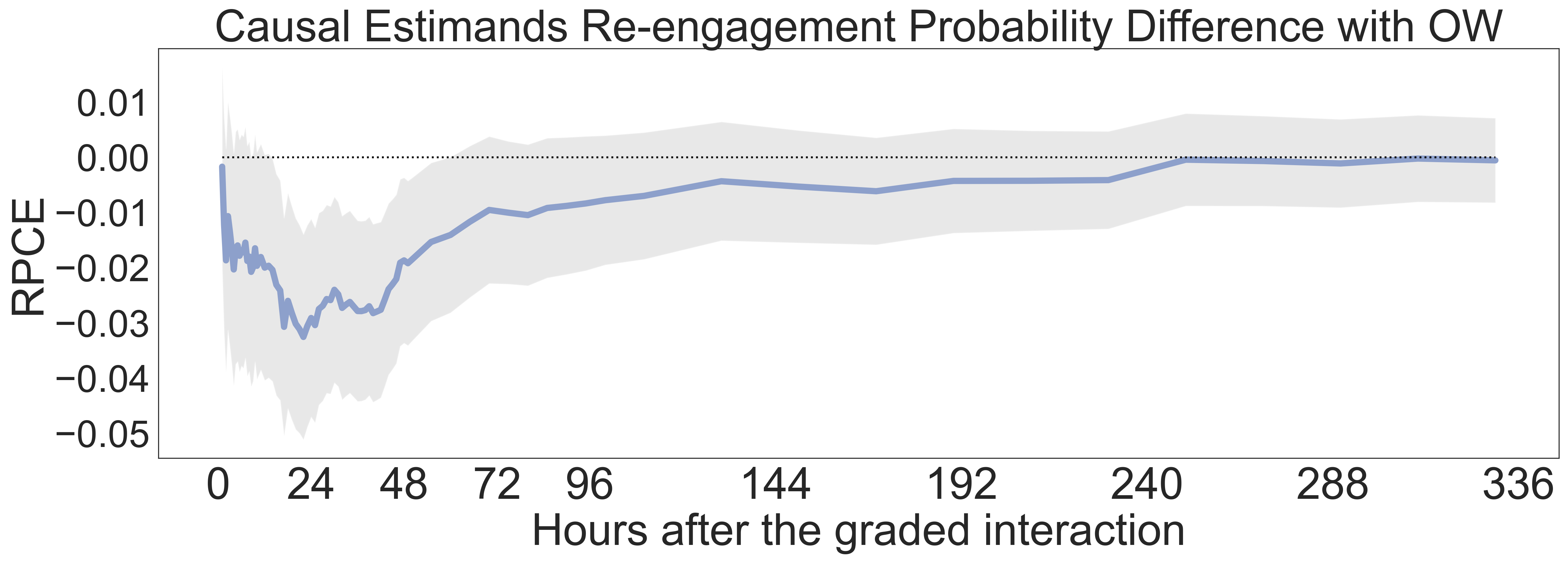}
    \vspace{-1.5em}
    \caption{The re-engagement probability causal estimands (RPCE) as a function of time after the annotated interaction, with associated 95\% confidence interval (shaded gray). 
    }
    \label{fig:time2next}
    \vspace{-2em}
\end{figure}

The estimator in Eq.\eqref{eq:pseudo-survival} represents the survival probability causal effect, \IE, the difference of the weighted \emph{re-engagement} probability in the \emph{Unhelpful} group and the \emph{Helpful} group, or the \emph{Re-engagement} Probability Causal Effect (RPCE). 
The results are shown in Figure~\ref{fig:time2next}. 
The confidence interval is calculated based on the estimated standard error of SPCE \cite{zeng2021propensity}.

The difference in estimated re-engagement probability is negative within the 336 hours (two weeks) following the initial interaction, with a maximum causal difference of $3.2\%$ at around 24 hours ($p$-value is $0.007$). The time window where the difference between the \emph{Helpful} and \emph{Unhelpful} groups is consistently statistically significant is between hours 8-65. The IPW result yields similar conclusions.
Our main takeaway is that the inhibition effect of an unhelpful interaction reaches peak around 24 hours after the interaction and then gradually weakening. 

Specifically, we conclude the following,
(i) An unhelpful interaction tends to have a stronger effect on whether the user wants to use the assistant again around the same hour on the next few days, perhaps affecting daily tasks like starting navigation to work,
(ii) About one week later, the re-engagement probability difference becomes insignificant, as users’ recollections of the unhelpful interaction fade away.

\vspace{-1.6em}
\subsection{Number of Active Days}\label{sec:NactiveDays}
\vspace{-0.8em}
Section \ref{sec:time2event} established the immediate effect of IA helpfulness on time-to-next engagement. In this section, we widen the analysis window and focus on the number of active days after the annotated interaction. Let $A^{(k)}$ denote the number of active days within $k$-day window, $k\in\{3,14\}$. The average treatment
effect (ATE) is defined as $\EE[A_i^{(k)}(1)-A_i^{(k)}(0)]$.

To estimate the causal effects with consistency, we applied four different statistical analysis tools at the two time windows respectively, belonging to two major branches of causal analysis. The first branch is weighting (IPW, entropy weights, overlap weights).\footnote{\tiny{Propensity weighting methods: https://cran.r-project.org/web/packages/PSweight}} The corresponding WATE function is defined similarly as Eq.\eqref{eq:WATE}.
The second branch is matching. Considering the dimensionality, we used the CEM method \cite{iacus2012causal}.


\begin{wrapfigure}{l}{.6\linewidth}
\vspace{-1em}
        \centering
    \includegraphics[width=\linewidth]{ 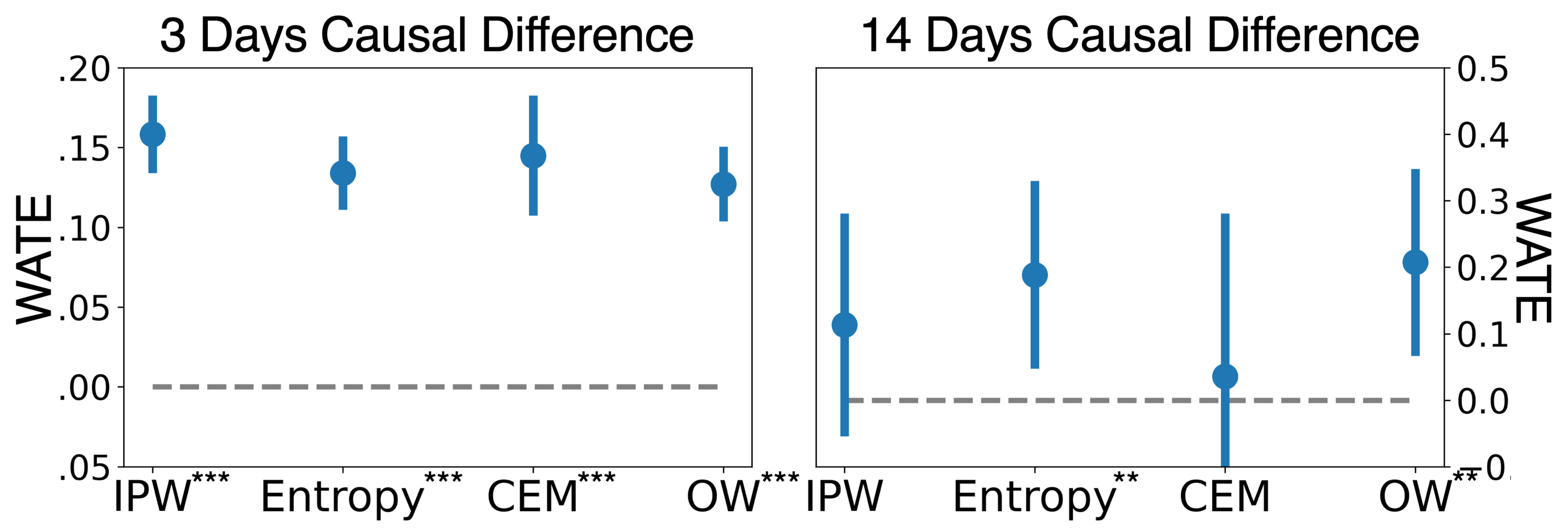}
    \vspace{-1.5em}
\caption{Causal effect of an unhelpful IA interaction on activity levels. Bar length indicates 95\% CI
}
\label{fig:N-Active-Days}
\vspace{-2em}
\end{wrapfigure}
In line with our previous findings, we observe statistically significant causal impacts on the activity level 3 days after the annotated IA interaction, shown in Figure~\ref{fig:N-Active-Days} (left). All four analysis tools yield $p$-values $<0.001$. 
This also supports the finding that the inhibition effect of an unhelpful engagement fades in time.
When we zoom out to a 14-day window, we observe that though the causal effects are not always significant, the directional consistency suggests a lessened effect of the unhelpful engagement compared to the 3-day window.




\vspace{-1.5em}
\section{Language Convergence in New User Cohort} \label{sec:semanticConvergence}

\vspace{-0.8em}


Having established the inhibition effect of an unhelpful interaction on a user's activity levels immediately following the interaction, we now expand both the scope and the time horizon of our analysis, to explore how prior engagements in turn shape users' linguistic choices over time. 

\vspace{-1.5em}
\subsection{New and Existing User Cohort Definition}
\vspace{-0.8em}
Canonically, a \emph{new user} to an IA is an individual who started using the IA for the first time in the observation window. As our data does not allow us to identify new users in this way, we
rely instead on the following conservative, \emph{necessary but insufficient}, condition for cohort determination: a user is assigned to the `new user cohort' if they (i) had at least one interaction with the IA in the study period, and (ii) had no interaction with the IA in the first 60 days of the study period. By erring on the side of including existing users in the new user group, we can
 ensure that any patterns that remain are robust. Therefore, we argue that this determination method offers a reasonable (and likely inflated) approximation of the true new user cohort. In our dataset containing 6 months of interaction data, approximately 17\% of all unique users were assigned to the new user cohort. 




\begin{table}[t]
    \centering
    \resizebox{.99\columnwidth}{!}{
\begin{tabular}{@{}lll@{}}
\toprule
\textbf{Low Perplexity}& \textbf{High Perplexity: syntactically complex sentences} & \textbf{High Perplexity: lexically diverse and rare topics}\\ \midrule
What is the \textbf{weather}, 3.3 & Show me hourly \textbf{weather} forecast, 17.1 & What is the \textbf{UV index}, 11.8  \\
 & Could I have the \textbf{weather} for rest of the week in \textless{}Location\textgreater please, 20.8 & Is there \textbf{tornado} nearby, 13.6\\
What is the \textbf{temperature}, 3.5 & When is the \textbf{rain} supposed to start again, 19.5 & How fast is the \textbf{wind} going, 15.6 \\
Will it \textbf{rain} today, 5.9 & When the \textbf{rain} going to stop, 21.8 & When is the full \textbf{moon}, 15.7 \\
Is it going to \textbf{snow} today, 7.6 & How many inches of \textbf{snow} are we supposed to get, 20.4 & What is the \textbf{barometric pressure} at \textless{}Location\textgreater{}, 27.1 \\
& How tall will the \textbf{snow} get tonight, 27.6 & \\ \bottomrule
\end{tabular}}
 \caption{Examples of low and high perplexity requests about weather.}\label{tab:PP_example}
\vspace{-3em}
\end{table}


%
%

%
\vspace{-1.5em}
\subsection{New user's self-Selection: Drop-out or adaption}
\vspace{-0.8em}
We use a domain-specific language model-based perplexity (PP) score \cite{lee2021towards}, which provides a comprehensive summary of the request's complexity characteristics \cite{adiwardana2020towards}.
PP score is defined as the inverse joint probability that a sentence belongs to the trained language model normalized by the number of tokens in the sentence \cite{jurafsky2000speech}, $PP(W) = \sqrt[N]{\frac{1}{\PP(w_1w_2...w_N)}}$,
where $W$ is the target sentence, $w_k$ is individual token and $N$ is the token count of the sentence. In our analysis, we adopted a tri-gram language model \cite{bird2009natural}.
Table \ref{tab:PP_example} presents examples of requests with perplexity scores.
Here we use paraphrased variants rather than actual user data for illustration purposes.
Higher perplexity correlates with more complex sentence structures,
more diverse language representations and broader topics.

\begin{figure}[ht!]
\vspace{-1.8em}
    \centering
    \includegraphics[width=.58\linewidth]{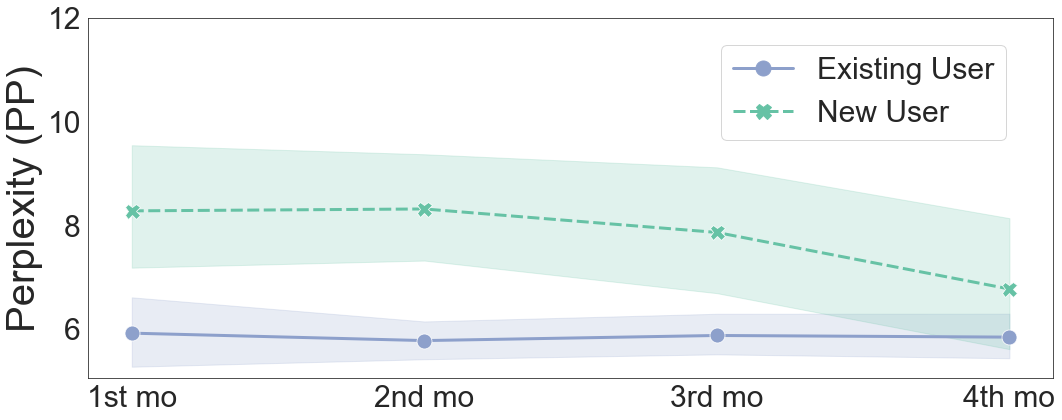}
    \vspace{-1.4em}
\caption{New user cohort has a higher average perplexity in the beginning and converges toward existing user cohort in language perplexity over time. \label{fig:convergence}
}
\vspace{-2.5em}
\end{figure}

Intuitively, new users tend to explore the limits of the IA system, with broader vocabulary and diverse paraphrases of their requests. 
In this study, we track the average PP scores of new and existing user cohorts over a six month period. 
First, we empirically show that the \emph{existing user} cohort has a lower and more stable perplexity score over time compared to the \emph{new user} cohort (Figure ~\ref{fig:convergence}). This result suggests that
requests from existing users are more likely to conform to the typical wording of requests within a certain domain.

Second, we discover that the perplexity score in the new user cohort is 30\% higher than in the existing user group in the first month, 
but it gradually converges to that of the existing user cohort.
This trend suggests that new users are less familiar with the IA's capabilities and are more exploratory when they are first introduced to the system. Over time, they gradually familiarize themselves with it. Eventually, they adopt similar sentence structures and other linguistic characteristics to those used by existing users when expressing similar intents. 

Next, within the new user cohort, we dive deeper into two subgroups:
The \emph{retained} group consists of users who were active for more than three out of the four month follow-up period. The  \emph{dropout} group includes users who were active for no more than 30 days within the study period.
Based on these criteria, the retained group has an average perplexity score of 7.5,
%
while the dropout group has an average perplexity score of 10.6 and the difference has $p$-value $<0.001$.
That is, users who stop using the IA within the first month tend to have substantially higher perplexity scores than users who are retained. Further, higher PP scores are closely related to higher unhelpful rates in the IA interactions. Following our findings from the previous sections, we expect unhelpful experiences to discourage users from continuing to engage with the IA.



In summary, we conclude that there are two plausible mechanisms that may explain the convergence of the perplexity score over time in the new user cohort:
\begin{enumerate}
    \item \textbf{Dropout}: some new users who are either unfamiliar with the supported functionality or the language of the IA system suffer negative experiences. These high-perplexity language users stop using the system after a few tries.
    \item  \textbf{Adaptation}: despite some potential negative experiences in the beginning, some new users familiarize themselves with the system and adapt to its limitations. They continue to use the system after the first few months.
\end{enumerate}

This represents a self-selection process among the users who choose to interact with the IA system: users adapt to the IA system in a way that lowers their language perplexity and consequently improves their experience, or they stop using it altogether.
Crucially, as users adapt their behavior to the system  over time, we expect to observe fewer and fewer requests that may lead to unhelpful interactions with the IA---as a result of the \emph{feedback effect}. Consequently, we observe a bias that introduces a significant challenge to the meaningful offline evaluation of the IA system based on naive samples of the usage traffic. 


\vspace{-1.2em}
\section{The Feedback Effect: Challenges to Meaningful Metrics}\label{sec:challenges}

\vspace{-0.3em}
\subsection{User-based vs. Usage-based evaluations}

Offline evaluation methodologies in the IA space mostly fall into two broad categories: \emph{user-based} and \emph{usage-based} approaches \cite{jiang2015automatic}. User-based approaches typically measure the overall satisfaction of a user, while usage-based approaches focus on success rates of the IA in correctly responding to a collection of requests. In this section, we discuss how the feedback effect introduces challenges to the construction of meaningful metrics for both types of approaches, informed by our findings in Sections \ref{sec:causal} and \ref{sec:semanticConvergence}.

\vspace{-0.8em}
\subsection{Implications of the Inhibition Effect}

As we established in Section \ref{sec:causal}, users who experience an unhelpful interaction with the IA are less likely to re-engage with it in the next few days. We may conclude, without loss of generality, that users who had unsatisfactory experiences in a preceding period are less likely to engage with the IA in the current period. Hence, if we are to survey \emph{active} users in a fixed time period to measure their expected satisfaction levels, unsatisfied users would have a lower probability of being surveyed than their satisfied counterparts. This ``heavy-user'' bias is ubiquitous in data mining \cite{wang2019heavy}. 


As an illustration, suppose that for the preceding period $T^{(k-1)}$ there were $N$ active users within the period, and no new users joined. Further, let $N'$ denote the number of re-engaged active users in the current period $T^{(k)}$ who were not active in $T^{(k-1)}$. Let $s$ be the share of users who were satisfied with their experiences with the IA, hence $(1-s)$ represents the unsatisfied share. Let $p$ be the probability of the satisfied users to re-engage in the current period (so that they may be surveyed), and $\Delta p$ indicates the difference in re-engagement probability for the unsatisfied users.
Among the re-engaged users in this period, there are $sN'$ satisfied users and $(1-s)N'$ unsatisfied users.

Accordingly, the total number of active users in the current period $T^{(k)}$ is $s p N + sN'$, and the total number of users remaining in the study is $s p N + (1-s)(p-\Delta p)N + N'$, and the estimand of user satisfaction rate in the current period $\hat{s}$ is defined accordingly. However, this is not an unbiased estimator of $s$ due to the feedback effect, shown in \eqref{eq:epsilon-s}. We further assume that the system of interest has reached long-term equilibrium s.t.\ the number of active users in adjacent time periods is nearly identical, and empirically $\Delta p$ is reasonably small s.t.\ $N'+p  N \simeq N$. With these assumptions, we propose an estimator of the measurement error as in \eqref{eq:epsilon-s-hat}.

\hspace*{-1.5em}
\begin{minipage}{0.59\linewidth}
\raggedleft
\begin{equation}\label{eq:epsilon-s}
\epsilon = \hat{s} - s = \frac{s \Delta p   (1-s)}{p-\Delta p + s  \Delta p + \frac{N'}{N}}    
\end{equation}
\end{minipage}
\begin{minipage}{0.42\linewidth}
\begin{equation}\label{eq:epsilon-s-hat}
\hat{\epsilon} = \frac{s\Delta p (1-s)}{1-\Delta p (1-s)}
\end{equation}
\end{minipage}
%
%

\begin{figure}[t]
    \centering
    \includegraphics[width=.8\linewidth]{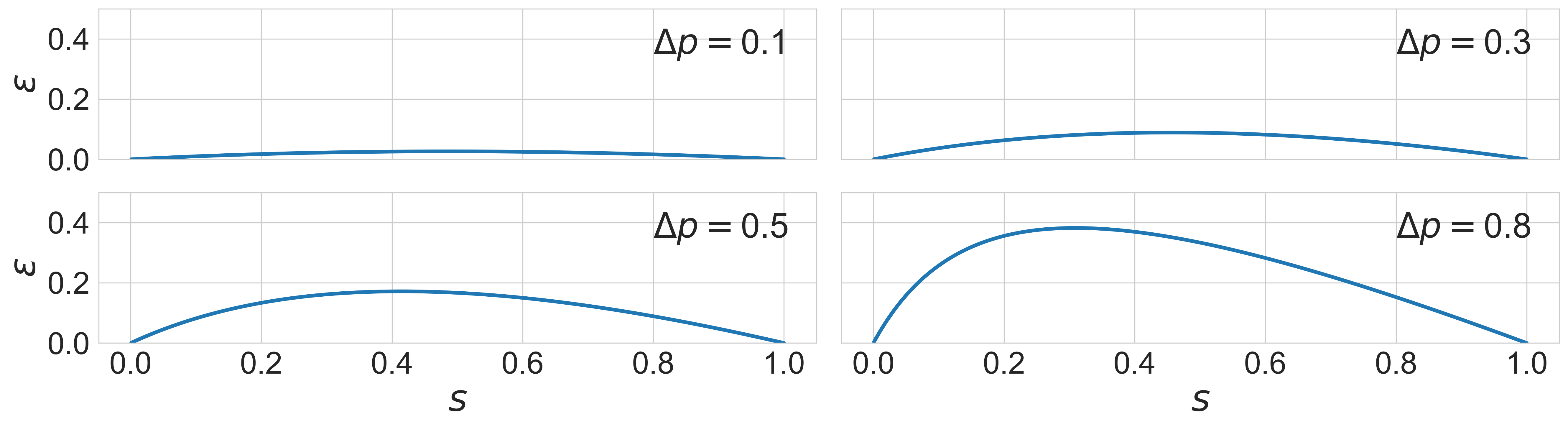}
    \caption{Estimated measurement error on user satisfaction rate for different $\Delta p$ based on \eqref{eq:epsilon-s-hat}}
    \label{fig:simulation}
    \vspace{-1.5em}
\end{figure}

For an IA with an actual user satisfaction rate $s=60\%$, $\Delta p = 0.3$, a simple survey of active users in the current period would yield a user satisfaction rate of $68\%$. A simulation study is presented in Figure~\ref{fig:simulation}. This error would be further amplified should the feedback effect (quantified by $\Delta p$) be stronger.

\vspace{-0.8em}


\subsection{Implications of the Language Convergence}

Language convergence has an equally profound impact on usage-based evaluation: as shown in  Table ~\ref{tab:helpfulness-preplexity}, the IA is nearly 3 times more likely to return a unhelpful response to high perplexity requests -- which account for a sizable share of the exploratory usage in the new user cohort -- than to low perplexity ones.

Should the true purpose of evaluation be to understand how well the IA addresses what the users \emph{truly want}, rather than a limited set of tasks that the users have \emph{compromised on}, we should always give sufficient consideration to the exploratory usage in our evaluation datasets. For emergent IAs with fast-growing user bases, this means one should carefully analyze the exploratory usage and iterate on new functionalities so that new users are retained at a higher level of perplexity, a reasonable proxy for request diversity and activity levels. For more established IAs, one should make a proactive effort to identify new user cohorts and decide on appropriate sampling strategies that balance the new and old.

\vspace{-0.8em}
\subsection{In Search of Meaningful Metrics: Some Recommendations}

While much of our discussion thus far has been in the context of IAs, it is perhaps not too wild a conjecture that the same phenomena can be observed more broadly in any intelligent systems that entail some form of an interactive user interface (e.g. dictation software, handwriting recognition, and etc.). We therefore provide some recommendations on how to construct more meaningful metrics in the presence the feedback effect:

\begin{enumerate}
    \item \textbf{Error Estimation and Selection Bias Mitigation}: for user-based studies, one may build an estimator to correct for measurement error after validating and quantifying the impact of the feedback effect on engagement; to control for the selection bias, one may elect to sample from a more comprehensive list of users (or the true population if feasible) rather than a list of active users in some fixed period to form the cohort of analysis;
    \item \textbf{Stratified Sampling along Multiple Dimensions}: one may identify key dimensions of interest to form stratified sampling strategies with enhanced coverage, such as system-designed function areas, user frequency, linguistic representations, and perplexities.
    \item \textbf{Exploratory Usage Retention:} exploratory usage often contains a more complete set of requests that users wish to accomplish through the system, and/or a more diverse set of user request patterns (accents in speech recognition, handwriting styles); it is a highly informative to collect data points that are yet to be subjugated to the inevitable influences of the feedback effect.
\end{enumerate}

\vspace{-1.3em}
\section{Discussions}\label{sec:discussion}
\vspace{-1em}
Evaluation of IA systems is an important yet challenging problem. On the one hand, the capabilities and limitations of IAs shape user behaviors (\EG, delayed engagement, dropout, and adaption). On the other hand, these very user behavior shifts in turn influence data collection and consequently the assessment of the IAs' capabilities and limitations. To our knowledge, this two-sided problem has not been formally discussed in the literature, at least in the context of real-world IAs. To fill this gap, this paper empirically studied the ``feedback effect'' nature of IA evaluation. On the one hand, we demonstrated that unhelpful interactions with the IA led to delayed and reduced user engagements, both short-term and mid-term. On the other hand, we examined long-term user behaviors, which suggested that as users gradually learned the limitations of the IA, they either dropped out or adapted (\IE, ``gave in''), and consequently increased the likelihood of helpful interactions with the IA.

Beside raising awareness within the data mining community, this paper aims to equip researchers and practitioners with tools for trustworthy IA evaluations. First, in cases where randomized controlled experiments are infeasible, we offered best practices on properly employing observational causal inference methods, and constructing offline metrics that take the censoring of user engagements into account. 
Second, to reduce the \emph{feedback loop} problem in data collection and sampling, it is important to gauge users' experience with the IA and control for confounding factors if possible. When not possible, researchers should consider stratified sampling or boosting the signals from more complex intents, or creating synthetic test data that varies in complexity, especially targeting more complex sentence structures and intent linguistic features which may be under-represented.
%
Third, we have demonstrated that a key factor contributing to unsatisfactory IA experiences for new users is that the language they use is too complex in some way. We have also shown that users who fail to adapt by using simpler language often do not continue to use the IA. These insights immediately suggest growth opportunities to capitalize on.  For example, multiple existing IAs offer a set of example conversations in different domains, in order to ``train'' new users to use the IA successfully right from the get go.

Our work implies multiple future directions, from both product and research perspectives. First, other than new user training (that might very well be skipped), what more can we do to convey the IA's capabilities and limitations, and help users engage more productively? Alternatively, how can we intervene early on and retain those ``drop-outs,'' who provide invaluable feedback to help improve our system? Second, although we collected a rich set of covariates to ensure unconfoundedness, we can further assess the robustness of the established causal links, by leveraging classic sensitivity analysis techniques \cite{liu2013introduction}. Third, while this paper focuses on off-line evaluation for IAs, it is possible to apply the proposed methodologies and recommendations in other settings (\EG, on-line experimentation) and software products (\EG, search engines). 

\vspace{-1.3em}
\subsection*{Acknowledgements}\label{sec:ack}
\vspace{-0.8em}
\small{
This work was made possible by Zak Aldeneh, Russ Webb, Barry Theobald, Patrick Miller, Julia Lin, Tony Y. Li, Leneve Gorbaty, Jessica Maria Echterhof and many others at Apple. We also thank Ricardo Henao and Shuxi Zeng at Duke University for their support and feedback.
}

\vspace{-1.5em}
\bibliographystyle{splncs04}
\small{\bibliography{FeedbackEffect-Full}}

\appendix
\beginsupplement
\beginsupplement

\appendix
\onecolumn %

\section*{Supplemental Material for ``How the Feedback Effect Shapes User Behavior with Intelligent Assistants''}
\section{Observational Study on IA's Helpfulness and users Engagement}
\subsection{Balancing Weights}
Li \citeSM{li2018balancing} proposed a family of balancing weights which enjoys balanced weighted distributions of covariates among treatment groups. , which enjoys balanced weighted distributions of covariates among treatment groups. Inverse-probability weights (IPW) is a special case of this family. Let $f(x)$ denotes the covariates distribution of the population, and $f_0(x), f_1(x)$ as the control or treatment group distribution respectively. 

With the balancing weights and tilting function $h(x)$, the weighted distributions for different treatment groups are evened out.
\begin{equation*}
    f_1(x)w_1(x)=f_0(x)w_0(x)=f(x)h(x),
\end{equation*}
The tilting function defines the target population and the estimands of interest, and also determined the weights accordingly. \begin{equation}\label{eq:balancingWt}
        \begin{cases}     w^{h}_1(x) \propto \frac{h(x)}{e(x)},& \text{for }Z=  1\\      w^{h}_0(x) \propto \frac{h(x)}{1-e(x)},& \text{for }Z=0. \end{cases}
\end{equation}
The population level causal estimands of interest, the weighted average treatment effect (WATE) shown in Eq.\eqref{eq:WATEfull}, is based on the balancing weights.
\begin{equation}\label{eq:WATEfull}
    \hat{\tau}=\hat{\mu}_1-\hat{\mu}_0=\frac{\sum_{i=1}^Nz^{h}_1(\bf{x_i})Z_iY_i}{\sum_{i=1}^Nz^{h}_1(\bf{x_i})Z_i} - \frac{\sum_{i=1}^Nz^{h}_0(\bf{x_i})Z_iY_i}{\sum_{i=1}^Nz^{h}_0(\bf{x_i})Z_i}.
\end{equation}
When $h(x)=1$, it is the inverse-probability weights (IPW). As the name suggested, the inverse of the probability that a unit is assigned to the observed group is the weight, and the corresponding estimand is ATE.

In Table~\ref{tab:balance}, it summarizes some weights from the balancing weights family.
\begin{table}[ht]
\centering
\begin{tabular}{@{}lll@{}}
\toprule
Method  & Target population & Tilting function $h(x)$                                                                                                              \\ \midrule
IPW     & Combined          & $1$                                                                                                                   \\
OW      & Overlapped        & $e(x)(1-e(x))$                                                                                                      \\
Entropy & Entropy based     & \begin{tabular}[c]{@{}l@{}} $e(x)\log(e(x))$ \\ - $(1-e(x))\log(1-e(x))$\end{tabular} \\ \bottomrule
\end{tabular}
\caption{Target population, tilting functions comparison for the balancing weights methods}\label{tab:balance}
\end{table}

\subsection{Outcome: time-to-next-engagement}
We have summarized our analysis pipeline in Algorithm~\ref{algo:causalSurvival}. Starting from estimating the pseudo-observations and propensity scores, then combining together to obtain the final weighted causal effect, re-engagement probability difference.
\begin{algorithm}[h]
\SetAlgoLined
{\bf Input:} Confounding variables $\Xmat$, IA helpfulness indicator $\Zmat$, observed time-to-next-engagement or censored time $\tilde{T}$ with censoring indicator $\Delta$.\\
\For{$t \in (0, t_{max}]$}{
{\bf Step 1:} Obtain pseudo-observations $\hat{\theta}_i(t)$\\
{\bf Step 2:} Estimate propensity score $\hat{e}(x)$\\
{\bf Step 3:} Calculate balancing weights $w_1(x), w_0(x)$ Eq.~\eqref{eq:OverlapWeight}\\
{\bf Step 4:} Derive the causal estimands of re-engagement probability difference (RPCE) $\hat{\tau}^w(t)$ Eq.~\eqref{eq:pseudo-survival}\\
}
 \caption{Causal discoveries of IA helpfulness on users time-to-next-engagement.}
 \label{algo:causalSurvival}
\end{algorithm}

\begin{figure}[ht]
    \centering
    \includegraphics[width=.8\linewidth]{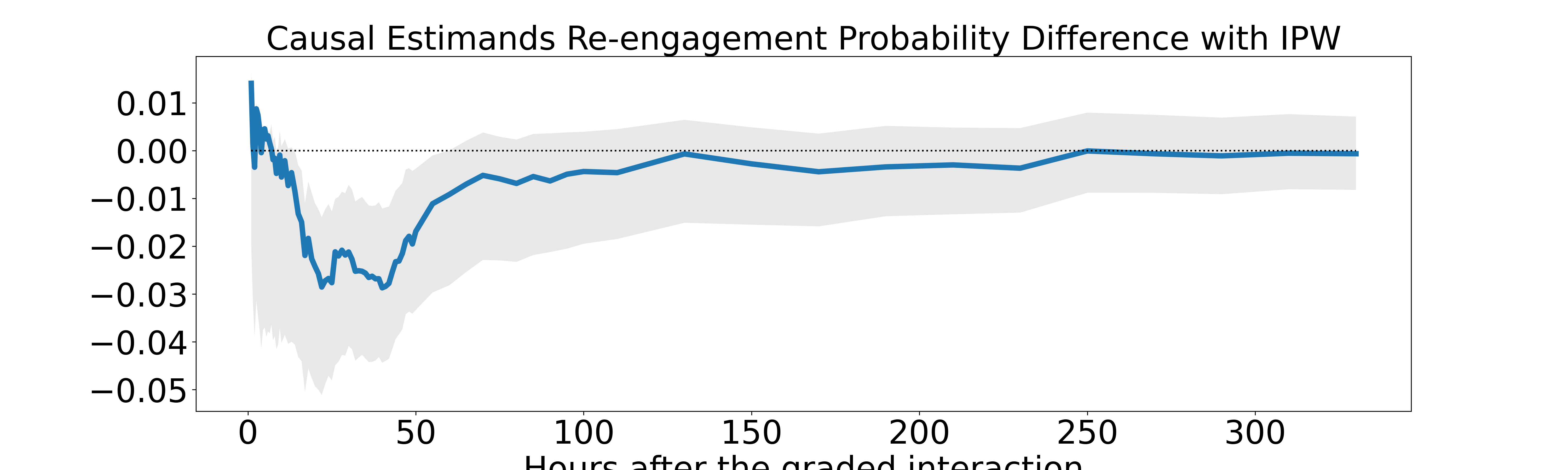}
    \caption{The re-engagement probability causal estimands (RPCE) as a function of time after the annotated interaction, with associated 95\% confidence interval (shaded gray). 
    The dotted horizontal line represents the difference is 0.
    The average re-engagement probability in \emph{Unhelpful} cohort is lower than \emph{Helpful} cohort in the following 336 hours (2 weeks) period. The significant gap happens around the $8 \sim 65$ hours with p-value $<0.05$.
    }
    \label{fig:IPWtime2next}
\end{figure}

Apart from the OW results presented in the main context, the IPW-weighted RPCE implies similar conclusions, as presented in Figure~\ref{fig:IPWtime2next}. The IPW results yield a significant difference in the time window between hours 16-50 with $p$-values $<0.05$. 

\section{Language Convergence}
\subsection{Discussions on Different Language Complexity Metrics}
Sentence complexity provides a linguistic measurement of user requests. Some simple metrics of sentence complexity are the number of tokens and the number of distinct N-grams per request \citeSM{xu2018diversity}. These are quite intuitive but hardly reveal deep insights into the requests \citeSM{evans2009myth}. To compare the sentence complexity of new requests with existing ones, there are three common metrics. Two of them, the selfBlEU score and the Jaccard similarity, are pairwise metrics based on overlapping N-grams between target and reference sentences \citeSM{zhu2018texygen,niwattanakul2013using}. The third one, word embedding diversity (WED) measures cosine distances of word embeddings of sentences in comparison.
Yet, these metrics are computationally inefficient owning to the enumeration of sentences in large datasets. On the other hand, the perplexity score (PP) is a scalable and informative metric that reveals both the syntactic complexity and the lexical diversity of a new request within a given text based on a trained language model \citeSM{holtzman2019curious}. Researches have also shown that the PP score of user requests inversely correlates with the success of the subsequent IA response: the higher the PP score, the more complex the request, the less likely the subsequent IA response is going to be successful \citeSM{adiwardana2020towards}.

\begin{figure}[ht]
    \centering
    \includegraphics[width=.7\linewidth]{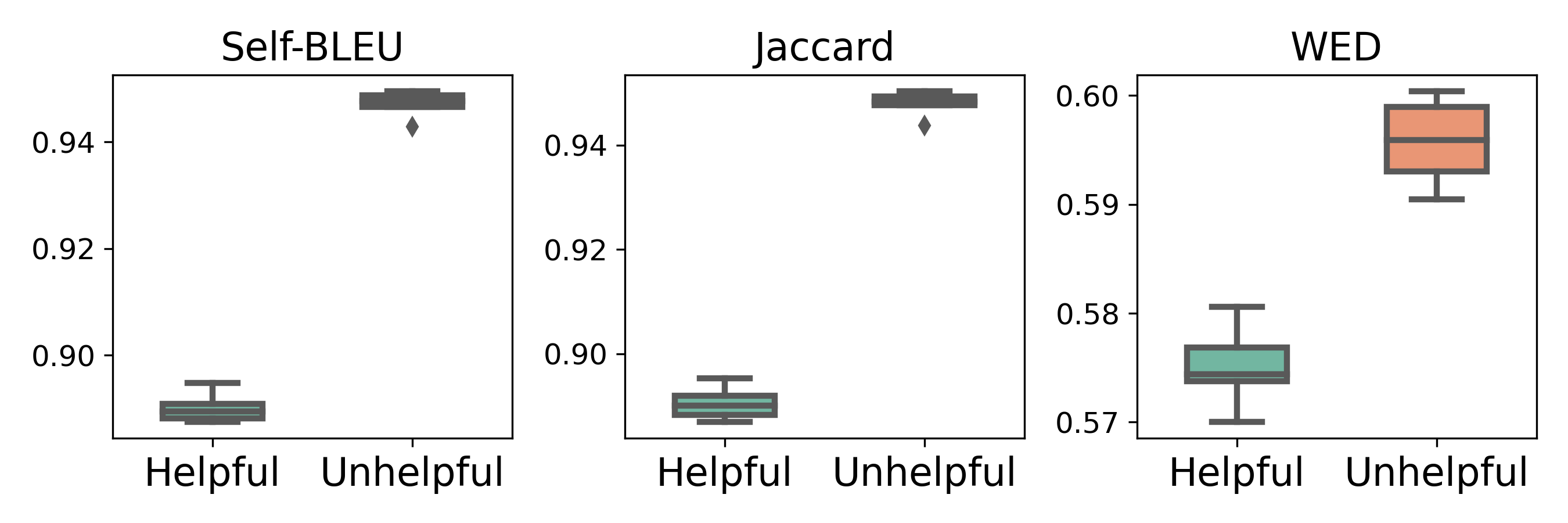}
    \caption{Language diversity for helpful utterances and unhelpful utterances. The bootstrapping results have shown that the unhelpful group has greater diversity with all three diversity metrics.}
    \label{fig:pairDiv}
\end{figure}

\begin{figure}[ht]
    \centering
    \includegraphics[width=.7\linewidth]{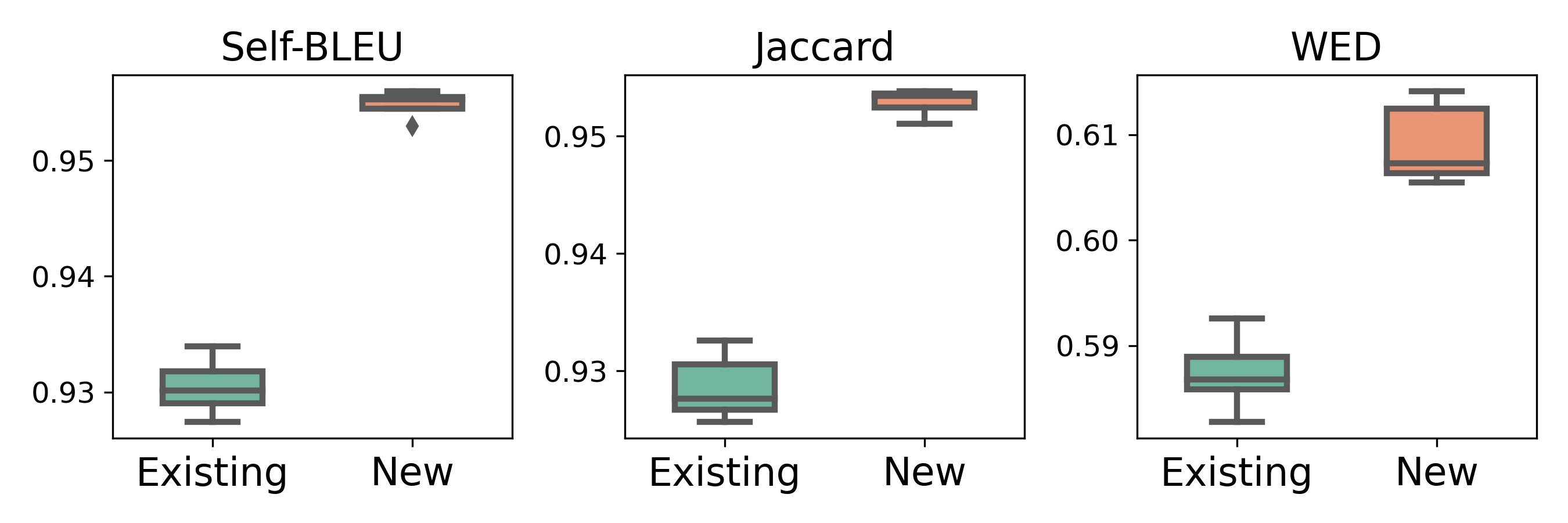}
    \caption{Language diversity for new and existing users cohort. Obviously the new group has greater diversity.}
    \label{fig:userDivPair}
\end{figure}

We first quantify the diversity with three widely used pairwise-based metrics without considering domain specific PP scores, i) selfBLEU: measures the diversity within the utterances cluster, i.e., average (1-pairwise BLEU score) ii) Jaccard similarity: average (1- pairwise word overlap) within the group iii) Word Embedding Diversity (WED): average (1- pairwise cosine distance) among embeddings of vectors in the utterances set. From Table~\ref{fig:pairDiv}, the unhelpful interactions have significantly larger diversities than the helpful group in general.
Also, to check the language diversity changing over time, we have run the above analysis on a group of likely new users (non-habitual) against a group of habitual users (in Figure ~\ref{fig:userDivPair}). From the exploratory analysis, the unhelpful interactions cohort and new users cohort share a high diversity in common.

 Revisiting the datasets described in Section~\ref{sec:causalData}, we further studied the correlation between language perplexity and the human-label review of helpfulness. 
As illustrated in the $2\times 2$ contingency table (Table ~\ref{tab:helpfulness-preplexity}), roughly 20\% of high perplexity requests are unhelpful, on the contrary, only 6\% of low perplexity ones are unhelpful. The association between perplexity score and IA helpfulness is statistically significant with p-value less than $0.0001$. 

\begin{table}[t]
\centering
\resizebox{.5\columnwidth}{!}{
\begin{tabular}{ccc|c}
\hline
          & High Perplexity & Low Perplexity & Total \\ \hline
Helpful   & 1245            & 11592          & 12837 \\
Unhelpful & 308             & 839            & 1147  \\ \hline
Total     & 1553            & 12431          & 13984 \\ \hline
\end{tabular}
}
    \caption{Contingency table of IA Quality (helpful vs.\ unhelpful) and PP score, with a statistically significant correlation. 
    }
    \label{tab:helpfulness-preplexity}
    \vspace{-2.5em}
\end{table}
%

Revisiting the datasets described in Section~\ref{sec:causalData}, we studied the correlation between language perplexity and the human-label review of helpfulness. 
About 20\% of high perplexity requests are unhelpful, compared to 6\% of low perplexity ones. 
The association between perplexity score and IA helpfulness is statistically significant with $p$-value $< 1\mathrm{e}{-4}$ with the Chi-squared test (Table ~\ref{tab:helpfulness-preplexity}).

\vspace{-0.8em}
\subsection{Evaluation Metrics for Request Complexity}

We evaluate user requests complexity along the following dimensions:
\begin{enumerate}
    \item \textbf{Syntactic complexity.} The same intent can be expressed in numerous ways, using varying levels of complexity of sentence structure. For example, ``call mom'' and ``place a telephone call to my mom please'' express the same intent, but the latter is more complex structurally. 
    \item  \textbf{Sub-intents entanglement within a request.} The level of detail required to address a user request may vary, with more complex intents requiring more detailed answers. For example, a simple request could be ``Is it going to snow today?'' and a more complex one could be ``How many inches of snow are we expecting over the next 7 days?''.
    \item  \textbf{Lexical and semantic diversity of the request.} Unlike common topics within a domain, infrequent topics are more difficult for the IA to handle. For example, within the weather domain, users commonly ask about the \emph{weather}, \emph{temperature}, and \emph{rain}, while diverse items include topics like \emph{wind}, \emph{tide}, \emph{barometric}, \emph{moon}, and \emph{tornado}. 
\end{enumerate}
%


%
Task-oriented IA systems often have a list of predefined domains: \textit{Weather}, \textit{Payment}, \textit{Phone}, \textit{Music}, \textit{LocalBusiness}, etc. Consequently, requests in a specific domain often share typical recurrent linguistic patterns. 
As a consequence, it is important to examine language diversity on a by-domain level, rather than on the dataset as a whole. This is because different domains may have different vocabulary sizes and high frequency tokens. 
To evaluate the complexity and diversity of a request, a simple metric like vocabulary size may be helpful in some utility domains (\EG, timer, alarm), but it could be misleading in communication domains (\EG,, phone call, SMS). Specifically, in communication domains, vocabulary size may be biased by the request's content (or payload) rather than the sophistication with which users express their intent. For example, if a user sends a longer message, the vocabulary size will increase accordingly, but the way they ask the IA to send the message may remain the same. 
Pairwise comparison metrics such as the previously mentioned selfBLEU, Jaccard, and WED are sensitive to keywords and topics, but they do not distinguish payload from non-payload content.

\break
\bibliographystyleSM{splncs04}
\bibliographySM{FeedbackEffect-Full}
\end{document}